\documentstyle[12pt]{article}

\advance\hoffset by -1.1truecm
\advance\voffset by -1.0truecm

\def\be{\begin{equation}}
\def\ee{\end{equation}}
\def\ba{\begin{eqnarray}}
\def\ea{\end{eqnarray}}

\def\bra#1{\langle #1 \rangle}

\begin{document}

\title{Uncertainty Relation in Quantum Mechanics
with Quantum Group Symmetry}
\author{Achim Kempf\thanks{supported by Studienstiftung des deutschen
Volkes, BASF-fellow}\\
Department of Applied Mathematics \& Theoretical Physics\\
University of Cambridge, Cambridge CB3 9EW, U.K.\\
 {\small Email: a.kempf@damtp.cambridge.ac.uk}}
\date{November 1993}
\maketitle

\begin{abstract}

We study the commutation relations, uncertainty
relations and spectra of  position and
momentum operators within the framework of quantum group
symmetric Heisenberg algebras
and their (Bargmann-) Fock representations.
As an effect of the underlying noncommutative
 geometry, a length and a momentum scale appear,
 leading to the existence of minimal nonzero
uncertainties in the positions and
 momenta. The usual quantum mechanical
behaviour is recovered  as a
 limiting case for not too small and not too
large distances and momenta.
 \end{abstract}

\vskip-14truecm
\hskip6.7truecm
\tt DAMPT/93-65 \quad and  \quad hep-th/9311147 \rm
\vskip14truecm

\section{Introduction}
The quantum group $SU_q(n)$ that we will use here
is a generalisation of the Lie group $SU_q(n)$. One recovers the
Lie group when the real
parameter $q$ tends to $1$. Technically we are dealing with
a non-commutative non-cocommutative quasitriangular Hopf algebra.
Its dual is a noncommutative generalisation of the
function algebra on the group manifold. We can thus consider
the quantum group as an example of noncommutative
 geometry \cite{DRI,FRT,MAJ1,Connes}.

It is tempting
to examine, whether noncommutative geometry,
when introduced into quantum theory,
regularises
its short distance behaviour. The idea is, not to \it
break \rm  symmetries
by this regularising procedure,
but to (quantum group- )  \it  generalise \rm   them instead.
One may then even speculate about gravity entering the picture.
This could be in such a
way that not only gravity is quantised but also that gravity
would feed back to quantum theory by modifying the canonical
commutation relations.

We will however for the present confine ourselves to the case
of nonrelativistic quantum mechanics.
 The study of
some effects of noncommutative
geometry in quantum mechanics was outlined in \cite{ixtapa}.
Here we cover a more general case
 and give details and proofs. Our results will support the
 idea that noncommutative geometry has indeed the potential to
 regularise ultraviolet and even infrared divergencies
 in quantum field theories.
\subsection{Heisenberg algebra}
In our approach we generate the Heisenberg algebra of $n$ degrees
of freedom by mutually adjoint operators $a_r$ and $a^{\dagger}_r$,
 $(r=1,...,n)$. This proceeding will automatically
supply us with a Hilbert (Fock-) space representation
of the Heisenberg algebra. In usual quantum mechanics this is
of course equivalent to the use of the hermitean generators $x_r$
and $p_r$, (which are the well known
linear combinations of the former ones) and the representation e.g.
on the Hilbert space of square integrable functions.

We will use the quantum group $SU_q(n)$ as a 'symmetry' group
for nontrivial commutation
relations i.e. as linear quantum canonical transformations.
Technically the Heisenberg algebra is a
Fun$SU_q(n)$-comodule algebra \cite{meinslmp}.
Arbitrary Hamiltonians can be studied within our framework
and they not necessarily have this symmetry.

Explicitely the commutation relations of the
following generalised bosonic Heisenberg algebra
 are conserved under the action of
 the quantum group $SU_q(n)$:
\begin{eqnarray}
a_i a_j - q a_j a_i & =  & 0 \mbox{ \quad for
\quad } i < j \label{eins}\\
a^{\dagger}_i a^{\dagger}_j -
q a^{\dagger}_j a^{\dagger}_i & =  & 0
\mbox{ \quad for \quad } i > j \label{zwei} \\
a_i a^{\dagger}_j  - q a^{\dagger}_j a_i & =  & 0
\mbox{ \quad for \quad } i \ne j \label{drei} \\
a_i a_i^{\dagger}
- q^2 a^{\dagger}_i a_i  & =  & 1
+ (q^2 - 1) \sum\limits_{j<i} a^{\dagger}_j a_j
\label{vier}
\end{eqnarray}
Here $i$ runs from $1$ to $n$ and $q$ is real.
These relations and their fermionic counterpart were derived
in the $R$-matrix approach in \cite{meinslmp}. As I learned later
they had first appeared in a different approach \cite{woro}. They
are related to the differential calculus on quantum planes
\cite{dcwess} and can also be understood as a braided
semi direct product construction \cite{MAJ2}. Compare
also with the different approaches
e.g. in \cite{macf,bieden,chef,arthur}.
Although quantum groups do in general
have more than one free parameter,
no further parameters enter in the above
commutation relations \cite{akny,berlin}.
\subsection{Bargmann Fock representation}
As usual the Fock space is constructed
 from a vector $\vert 0\rangle$ with
$$ \bra{ 0 \vert 0 } = 1 \mbox{\quad
and\quad } {a_i \vert 0 \rangle } = 0
\mbox{\quad for\quad } i = 1,...,n $$
One then obtains for the scalar product:
\begin{equation}
\langle 0\vert (a_n)^{r_n} \cdot ... \cdot (a_1)^{r_1}
 (a_1^{\dagger})^{r_1} \cdot ... \cdot
 (a_n^{\dagger})^{r_n} \vert 0\rangle \quad  =
 \prod\limits_{i=1}^n [r_i]_{q}!
\end{equation}
with
\begin{equation}
[r]_q! := [1]_q \cdot [2]_q \cdot [3]_q
\cdot . . . \cdot [r]_q \mbox{ \qquad and \qquad }
[p]_q := \frac{q^{2p} - 1}{q^{2} - 1}
\label{norm}
\end{equation}
Thus the scalar product remains positive definite.
The Hilbert space $\cal{H}$, completed
 using the induced norm, is as usual
isomorphic to the space of square summable series $l^2$.
The Poincar{\'e} series of the $a$'s and the
$a^{\dagger}$'s remain unchanged.
\medskip\newline
The usual
quantum mechanical programme proved to be possible:
The Heisenberg algebra can be represented
on a positive definite
(Bargmann Fock-) Hilbert space of
 wave functions and integral
 kernels like e.g. Green functions can be defined.
The one dimensional case had early been studied in
\cite{baulieu}. A new 'integral' for
calculating the scalar product of two
Bargmann Fock wave functions was developed
\cite{meinslmp,meinsjmp}, having the
same\footnote{unlike using as usual ordinary complex integration for
the bosonic case and Berezin integration for the fermionic case}
evaluation procedure in the bosonic as in the fermionic case.
Using the undeformed Schr{\"o}dinger equation,
i.e. $i\hbar\partial_t \Psi = H \Psi$ with $\partial_t$ an
ordinary derivative,
some dynamical systems were worked out and lead
 to unitary time evolution \cite{meinsjmp}.

Since the above given commutation relations are
respected by formal hermitean conjugation, the
natural candidates for position and
momentum operators
$ x \propto a + a^{\dagger}$ and
$p \propto i(a^{\dagger}-a)$
are representable as symmetric operators
on a suitable domain. Let us now
try to reveal some features of the
 underlying noncommutative geometry by
studying these observables in detail.

\section{Position and momentum operators}
We start with the following ansatz
 for the position and momentum operators:\newline
$(r = 1, ... ,n)$
\begin{equation}
x_r := L_r (a^{\dagger}_r + a_r) \mbox{ \qquad and \qquad }
p_r := i K_r (a^{\dagger}_r - a_r)
\label{xpansatz}
\end{equation}
Defining their domain $D$ to be
\begin{equation}
D := \{ v \in {\cal{H}} \vert  v =
\mbox{Polynomial}(a^{\dagger}_1,...,a^{\dagger}_n)
\vert 0\rangle \}
\end{equation}
which is dense in $\cal{H}$,
 we insure that all $x_r$ and $p_r$ are
 represented as symmetric operators
 with images that lie in their domain.
Since the $a$'s and $a^{\dagger}$'s
 do not carry units, the newly
introduced constants $L$ and $K$ do.

It is reasonable
to require the existence of a physical region in which
the usual quantum mechanics is recovered as a limiting
 case\footnote{weakening this restriction, one may
 generalisable the ansatz Eqs.\ref{xpansatz} }, even if $q^2 \ne 1$.
This could be achieved if the commutation
relations come out in the form
$[ x,p ] = i \hbar + f(q,x,p)$
i.e. with the central term being $i\hbar$
without any $q$- factors. The uncertainty relation
$\Delta x \Delta p \ge \frac{\hbar}{2} + \frac{1}{2}
 \langle f(q,x,p)\rangle $
should then reduce to the usual relations where
$\langle f \rangle $ is negligible. The actual commutation relations
come out as follows:
\subsection{Commutation relations}
We express the commutation relations Eqs.\ref{vier}
 in terms of the $x$'s
and $p$'s:
\begin{eqnarray}
[x_r,p_r] & = &  \frac{ 4 i L_r K_r}{q^2 +1}\\
          &   & + \frac{4 i L_r K_r (q^2-1)}{q^2+1}
\left[ \sum_{s\le r}
\left( \frac{x_s^2}{4 L_s^2}+\frac{p_s^2}{4 K_s^2} \right)
 - \sum_{t < r} \frac{1}{4 i L_t K_t} [x_t,p_t] \right]\nonumber
\end{eqnarray}
The commutators $[x_t,p_t]$ (with $t<r$) can be eliminated by
iterating the equation. Choosing the products $L_s K_s$ appropriately
we can indeed bring the commutation relation into the desired form,
i.e. with the central term being exactly $i\hbar$:
\smallskip\newline
The ansatz
\be
[x_r,p_r] = i\hbar +i  \sum_{s\le r} {\beta}_s
\left(\frac{x_s^2}{4 L_s^2}+\frac{p_s^2}{4 K_s^2} \right)
\ee
leads for the central term to the equation:
\be
i \hbar = \frac{ 4 i L_r K_r}{q^2 +1}
- i \hbar L_r K_r \frac{q^2-1}{q^2+1} \sum_{t < r} \frac{1}{L_t K_t}
\ee
Solving this equation we get relations between the $L_r$ and $K_r$:
\begin{equation}
L_r K_r := \frac{\hbar}{2} \left(\frac{q^2 +1}{2}\right)^r
\label{klpbez}
\end{equation}
{}From this follows ${\beta}_r$ immediately:
\be
{\beta}_r = 4 L_r K_r \frac{q^2-1}{q^2+1} =
\hbar (q^2-1)  \left(\frac{q^2 +1}{2}\right)^{r-1}
\ee
Thus the commutation relations take final form:
\begin{equation}
[x_r,p_r] = i \hbar + i \hbar (q^2 -1)
\sum_{s\le r}
\left( \frac{q^2+1}{2}\right)^{s-1}
\left(\frac{x_s^2}{4 L_s^2}+\frac{p_s^2}{4 K_s^2} \right)
\label{xpcr}
\end{equation}
The mixed commutation relations read for $s>r$:
\begin{equation}
[x_s,p_r] =  i \frac{L_r}{K_r} \frac{q-1}{q+1} \{ x_s, x_r\}
\label{bspder}
\end{equation}
\begin{equation}
[x_s,x_r] =  i \frac{K_r}{L_r} \frac{q-1}{q+1} \{ x_s, p_r\}
\label{pospos}
\end{equation}
For $s<r$ one gets:
\begin{equation}
[x_s,p_r]  =  i \frac{K_s}{L_s} \frac{q-1}{q+1} \{ p_s, p_r\}
\end{equation}
\begin{equation}
[p_s,p_r]  =  -i \frac{L_s}{K_s} \frac{q-1}{q+1} \{ x_s, p_r\}
\end{equation}
To see this, solve Eqs.\ref{xpansatz} for the $a$'s
and $a^{\dagger}$'s,
express Eqs.\ref{eins}-\ref{drei} in terms of the $x$'s and $p$'s
and find e.g. Eq.\ref{bspder} from Eq.\ref{eins} $+$
(Eq.\ref{eins})$^+$
 $+$ Eq.\ref{drei} $+$ (Eq.\ref{drei})$^+$.
If $q^2 =1$ the constants $K$ and $L$ drop out of
 the commutation relations, reflecting
that in ordinary quantum mechanics a length or
 a momentum scale can only be set
by the Hamiltonian i.e. by choosing a particular system.
 Here, for $q^2 \ne 1$ the
$K$ and $L$ appear in the commutation relations,
 thus these scales become a property
of the quantum mechanical formalism itself.
\subsection{A maximal set of commuting observables}
The operators $x_i$ (as well as the operators $p_i$) no longer
commute among themselves. Thus we conclude that the position
operators can not be simultaneously diagonalised and the same for
the momentum operators.

Before studying the structure of the noncommutative
'configuration space' and the
 noncommutative momentum space in more
detail, let us mention that for
 example the following symmetric operators
$h_i$ can still serve as a set of commuting observables:
\be
h_i :=
\frac{x_i^2}{4 L_i^2} + \frac{p_i^2}{4 K_i^2}
\qquad \quad (r=1,...,n)
\ee
The symmetry is obvious. To prove commutativity we use that
\be
[a^{\dagger}_i a_i, a^{\dagger}_j a_j] =
0 \qquad \quad \forall i,j
\ee
which follows immediately from Eqs.\ref{eins}-\ref{drei}.
\smallskip\newline
With the definitions Eqs.\ref{xpansatz} follows
\be
a^{\dagger}_r a_r =
\frac{x_r^2}{4 L_r^2} + \frac{p_r^2}{4 K_r^2}
 - \frac{1}{4 i L_r K_r} [x_r,p_r]
\ee
Eliminating the commutators using Eq.\ref{xpcr} yields
\be
a^{\dagger}_r a_r =
h_r - \frac{1}{4 i L_r K_r}
\left( i\hbar + i\hbar (q^2 -1)
\sum_{s\le r}
\left( \frac{q^2+1}{2}\right)^{s-1} h_s \right)
\ee
which eventually gives
\be
[h_i,h_j] = 0 \qquad \quad \forall i,j
\ee
The eigenstates of the operators $h_i$ are of course generalised
Hermite wave functions. Let us however focus on the
operators $x$ and $p$, i.e on configuration and momentum space.
\section{Uncertainty relation}
For simplicity we first consider  the
 $1$ dimensional case where Eq.\ref{xpcr}
reads:
\begin{equation}
[x,p] = i \hbar + i \hbar (q^2-1)
 \left(\frac{x^2}{4L^2} + \frac{p^2}{4K^2}
\right)
\label{1dimcr}
\end{equation}
with
\begin{equation}
K = \frac{\hbar}{4L} (q^2+1)
\label{klbez}
\end{equation}
We will now study the situation for $q^2 >1$.
The case $q^2<1$ is quite
different and will be discussed elsewhere.
The following (standard) derivation of the uncertainty relation
holds on every
domain $D^{\prime}$ of $x$ and $p$,
 on which both operators are symmetric and
have their images in the domain.
 The above given domain $D$ is an example.
\medskip\newline
We start with the trivial statement that
 the following norm is positive:
$$
\vert \left( (x-\langle v,x.v\rangle )
 +i\alpha (p-\langle v,p.v\rangle )\right) v \vert \ge 0
\mbox{\qquad \quad} \forall v \in D^{\prime}
\mbox{\quad } \forall \alpha $$
Using that $x$ and $p$ are symmetric on $D^{\prime}$ this is for all
real $\alpha $:
\be
(\Delta x)^2 + \alpha ^2 (\Delta p)^2 + i \alpha
\bra{v,[x,p].v} \ge 0
\ee
with the usual definitions, e.g.:
\be
(\Delta x)^2 := \bra{v,(x - \bra{v,x.v})^2.v}
\ee
This can be put into the form:
\be
(\Delta p)^2 \left( \alpha - \frac{\hbar A}{\alpha
(\Delta p)^2 }\right)^2
- \frac{{\hbar}^2 A^2}{4 (\Delta p)^2 } + (\Delta x)^2 \ge 0
\label{vucr}
\ee
with $A$ coming from the rhs of the
commutation relation Eq.\ref{1dimcr}:
\be
A =  1 + (q^2-1)\left(
\frac{(\Delta x)^2 + \langle x\rangle ^2}{4L^2}
 + \frac{(\Delta p)^2 + \langle p\rangle ^2}{4K^2}\right)
\ee
(we abbreviated e.g. $\bra{v,x.v}$ as $\bra{x}$)\newline
Now for any given $v \in D^{\prime}$ we get the most restrictive
inequality when choosing $\alpha = \hbar A/2 (\Delta p)^2$ so
 that the first term on the lhs of Eq.\ref{vucr},
 which can't be negative, vanishes. This
yields for all $v$ in $D^{\prime}$ the uncertainty relation
\begin{equation}
\Delta x \Delta p \ge \frac{\hbar}{2} \left( 1 + (q^2-1)\left(
\frac{(\Delta x)^2 + \langle x\rangle ^2}{4L^2}
 + \frac{(\Delta p)^2 + \langle p\rangle ^2}{4K^2}\right)
\right)
\label{ucr1}
\end{equation}
\subsection{Minimal uncertainties in position and momentum}
In order to analyse the content of this
uncertainty relation we express it in 'polar coordinates'
\begin{equation}
\Delta x := 2 L r \cos\alpha \mbox{ \quad and \quad }
 \Delta p := 2Kr \sin\alpha
\end{equation}
where it takes the form:
\begin{equation}
r^2 \ge \frac{ 1 + (q^2-1)
\left( \frac{\langle x\rangle ^2}{4L^2} +
 \frac{\langle p\rangle ^2}{4K^2}\right)}
{(q^2+1) \sin2\alpha -  (q^2-1)}
\label{pol2}
\end{equation}
with
\begin{equation}
\sin2\alpha > \frac{q^2-1}{q^2+1}
\label{pol1}
\end{equation}
{}From Eq.\ref{pol1} follows that the
minimal $\alpha$ is larger than $0$ and
the maximal $\alpha$ is smaller than $\pi/2$. Thus the
hyperbola of the ordinary uncertainty relation,
 having the $\Delta x$ and the $\Delta p$
axes as asymptotes has turned into a graph
with asymptotes that are no longer
parallel to the axes.
{}From Eq.\ref{pol2} follows
that $r$ is always larger than $0$. We thus conclude that
there are minimal nonzero
uncertainties in the positions and the momenta.
\smallskip\newline
In order to calculate them we define
\be
f(\Delta x,\Delta p) :=
\Delta x \Delta p - \frac{\hbar}{2} \left( 1 + (q^2-1)\left(
\frac{(\Delta x)^2 + \langle x\rangle ^2}{4L^2}
 + \frac{(\Delta p)^2 + \langle p\rangle ^2}{4K^2}\right)
\right)
\ee
and find e.g. $\Delta x_{min}$ by solving
\be
\frac{\partial }{\partial \Delta p} f(\Delta x,\Delta p) = 0
\mbox{\qquad and \qquad} f(\Delta x, \Delta p) = 0
\ee
which has the solution:
\be
(\Delta x_{min})^2 = L^2 \frac{q^2-1}{q^2}
 \left( 1 + (q^2-1)\left(
\frac{\langle x\rangle ^2}{4L^2}
 + \frac{\langle p\rangle ^2}{4 K^2}\right)\right)
\ee
Thus the absolutely smallest uncertainty in the position is:
\begin{equation}
\Delta x_{0} = L \sqrt{\frac{q^2-1}{q^2}}
\ee
Analogously one obtains the absolutely smallest uncertainty in the
momentum:
\be
\Delta p_{0} = K \sqrt{\frac{q^2-1}{q^2}}
\end{equation}
\newline
%
%
%
%
%
%
Due to Eq.\ref{klbez} there were two
free parameters: The length $L$ and $q$.
Instead we can now use $\Delta x_{0}$
and $\Delta p_{0}$ as the free
parameters and express $L$,$K$ and $q$ in terms of these:
\begin{equation}
L= \Delta x_{0} \sqrt{\frac{2 \Delta x_{0} \Delta p_{0}
  + \hbar + \sqrt{4 (\Delta x_{0})^2
(\Delta p_{0})^2 + (\hbar)^2}}{
 4 \Delta x_{0} \Delta p_{0}}}
\end{equation}
\begin{equation}
K= \Delta p_{0} \sqrt{\frac{2 \Delta x_{0} \Delta p_{0}
  + \hbar + \sqrt{4 (\Delta x_{0})^2
 (\Delta p_{0})^2 + (\hbar)^2}}{
 4 \Delta x_{0} \Delta p_{0}}}
\end{equation}
\begin{equation}
q = \sqrt{\left( 2\Delta x_{0} \Delta p_{0} +
\sqrt{4 (\Delta x_{0})^2
(\Delta p_{0})^2 + {\hbar}^2 }\right)/\hbar }
\end{equation}
The commutation relation Eq.\ref{1dimcr} then takes the form:
\begin{equation}
[x,p] = i\hbar + i g( \Delta x_{0}, \Delta p_{0})
\left( \frac{x^2}{(\Delta x_{0})^2}
+ \frac{p^2}{(\Delta p_{0})^2}
\right)
\end{equation}
where
\begin{equation}
g(\Delta x_{0}, \Delta p_{0}) :=
 4 \frac{\Delta x_{0} \Delta p_{0}}{\hbar}
\frac{2 \Delta x_{0} \Delta p_{0}
 + \sqrt{4 (\Delta x_{0} \Delta p_{0}
)^2 +{\hbar}^2} -\hbar}{2 \Delta x_{0} \Delta p_{0} +
\sqrt{4 (\Delta x_{0} \Delta p_{0}
)^2 +{\hbar}^2} +\hbar}
\end{equation}
Let us now identify the physical
 region where the ordinary quantum
mechanical behaviour is recovered:

Since physically we know that $\Delta x_{0}$
and $\Delta p_{0}$ can only be very small, say
$\Delta x_{0} \Delta p_{0} \ll \hbar /2$,
 we expand $g$ to the first
nonzero order in this product
and arrive at the simplified commutation relation:
\begin{equation}
[x,p] = i\hbar + \frac{4i}{\hbar} \left( x^2 (\Delta p_{0})^2 +
p^2 (\Delta x_{0})^2 \right)
\end{equation}
Now it becomes clear how in our formalism
not only the behaviour for small distances
and momenta is altered: Also
 for expectation values of $x^2$ or $p^2$
large enough to make the second term on the
 rhs of the order $\hbar$ or larger,
the behaviour will be significantly changed.
 The region of approximately
ordinary quantum mechanical behaviour is thus specified through:
\begin{equation}
 (\Delta x_{0})^2 \ll x^2 \ll
 \frac{{\hbar}^2}{4(\Delta p_{0})^2}
\end{equation}
\begin{equation}
 (\Delta p_{0})^2 \ll p^2 \ll \frac{{\hbar}^2}{4(\Delta x_{0})^2}
\end{equation}
{}From the point of view of wave-particle dualism, meaning e.g. that high
momenta are needed to measure small
distances, this is of course a
reasonable result.
\subsection{Generalisation to $n$ dimensions}
Let us now return to the $n$ dimensional case where we will recover
essentially the same behaviour:

It is straightforward to see that the derivation of Eq.\ref{ucr1}
also works in the $n$ dimensional case i.e. we get:
\be
\Delta x_j \Delta p_j \ge \frac{1}{2} \vert \bra{[x_j,p_j]}\vert
\ee
where the commutator is given in Eq.\ref{xpcr}. There are again
minimal uncertainties $\Delta x_{i_{min}}$ and $\Delta p_{i_{min}}$
which are calculated using
\be
f(\Delta x_j, \Delta p_j) := \Delta x_j \Delta p_j
- \frac{1}{2} \vert \bra{[x_j,p_j]}\vert
\ee
and solving, in order to obtain e.g. $\Delta x_{j_{min}}$:
\be
\frac{\partial }{\partial \Delta p_j} f(\Delta x_j, \Delta p_j)
\mbox{ \qquad and \qquad }
f(\Delta x_j, \Delta p_j) = 0
\ee
The solutions are:
\be
\Delta x_{j_{min}} = L_j \frac{q^2 -1}{q^2} \left(
\frac{q^2+1}{2}\right)^{1-j} \cdot s(j)
\mbox{ \quad and \quad }
\Delta p_{j_{min}} = K_j \frac{q^2 -1}{q^2} \left(
\frac{q^2+1}{2}\right)^{1-j} \cdot s(j)
\ee
with
\begin{eqnarray}
s(j) = 1  & + & (q^2-1) \sum_{s<j}
 \left( \frac{q^2+1}{2}\right)^{s-1}
\left( \frac{\bra{x^2_s}}{4 L_s^2} +
 \frac{\bra{p^2_s}}{4 K_s^2}\right)
\nonumber \\
  &   & + (q^2 -1)  \left( \frac{q^2+1}{2}\right)^{j-1}
\left( \frac{\bra{x_j}^2}{4 L_j^2} +
 \frac{\bra{p_j}^2}{4 K_j^2}\right)
\end{eqnarray}
Due to Eq.\ref{klpbez} the $K_j$ and
 the $L_j$ are not independed, so that
we are left with in total $n +1$ free parameters.
 One may choose them
to be e.g. $q$ and the $n$ minimal uncertainties in the positions.

It would now of course be interesting to study whether symmetric
operators of angular momentum can be expressed in terms of the
position and momentum operators and how they reflect the
appearance of minimal uncertainties. Let us however first study the
functional analysis of the position and momentum operators in more
detail.
\section{Functional analysis of $x$ and $p$}
We first consider the one dimensional case.
The above derived uncertainty relation Eq.\ref{ucr1}
 holds on every domain $D^{\prime}$
on which both, $x$ and $p$ are symmetric
and have their ranges in $D^{\prime}$.
The uncertainty relation
implied minimal nonzero uncertainties in the positions and momenta.
\newline
Now
if there was a $v_{\lambda} \in D^{\prime}$
that is eigenvector e.g. of
$x$: \quad
\be x.v_{\lambda} = \lambda v_{\lambda}
\ee
one would then of course have
\be
(\Delta x)^2 = \langle v_{\lambda}\vert (x-\langle
 v_{\lambda},x.v_{\lambda}\rangle )^2\vert
v_{\lambda}\rangle = 0
\ee
which would be a contradiction.
We thus conclude that there is no domain on which $x$ and $p$ are
symmetric {\it and \rm} have eigenvectors.
Let us now study the functional analysis of $x$ in more detail,
the analysis for $p$ is completely analogous.
\subsection{The operators $x$, $x^*$ and $x^{**}$}
We start be choosing for $x$ the domain $D_x := D$
 (the finite linear combinations
of the vectors $(a^{\dagger})^r
\vert 0\rangle $ with $r = 0,1,2,...$),
 on which $x$ and $p$ are obviously symmetric
 and have their image in $D_x$.
We can thus already conclude from above
that $x$ has no eigenvectors in $D_x$.
Indeed, the eigenvalue problem
\begin{equation}
 x.v_{\lambda} = \lambda  v_{\lambda}
\mbox{ \qquad with \qquad }
v_{\lambda} = \sum_{r=0}^{\infty} f_r(\lambda )
 \frac{(a^{\dagger})^r}{
\sqrt{[r]_q!}} \vert 0\rangle
\end{equation}
can be solved for all complex $\lambda$,
but from the recursion formula
that we obtain for the coefficients
$f_r(\lambda )$ of $v_{\lambda}$ it is clear that
 infinitely many of them are nonzero, thus $v_{\lambda} \not\in D_x$:
Explicitely, in the orthonormalised basis
\be
e_r :=  \frac{(a^{\dagger})^r}{\sqrt{[r]_q!}} \vert 0\rangle
\ee
the matrix elements of $x$ are
\be
x_{rs} = L (\sqrt{[r]_q} {\delta}_{r,s+1} + \sqrt{[s]_q}
{\delta}_{r+1,s})
\ee
and the recursion formula that we obtain for the coefficients
$f_r(\lambda)$ of the vector $v_{\lambda}$ thus reads:
\be
\sqrt{[r+1]_q} f_{r+1} = \frac{\lambda}{L} f_r -
\sqrt{[r]_q} f_{r-1}
\ee
Let us now consider the adjoint $x^*$ of $x$. It is a
closed operator since ${\bar{D}}_x = \cal{H}$ and has the domain:
\begin{equation}
D_{x^{*}} = \left\{ v \in {\cal{H}} \vert \quad \exists w \in
{\cal{H}} \quad \forall a \in D_{x} :\quad
  \langle v,x.a\rangle  = \langle w,a\rangle \right\}
\end{equation}
Of course $D_x \subset D_{x^*}$ and,
using the above mentioned recursion formula,
one proves that actually all $v_{\lambda}$
are normalisable and are contained
in the domain $D_{x^*}$:
\newline
To see this we rewrite the recursion formula in matrix form:
\smallskip\newline
\be
\left( \matrix{f_{r+1} \cr \cr f_r}\right) =
\frac{1}{\sqrt{[r]_q[r+1]_q}}
\left( \matrix{\frac{{\lambda}^2}{L^2} -[r]_q, &
-\frac{\lambda}{L} \sqrt{[r-1]_q} \cr \cr
\frac{\lambda}{L} \sqrt{[r+1]_q},
& -\sqrt{[r-1]_q [r+1]_q }} \right)
\left( \matrix{f_{r-1} \cr \cr f_{r-2}} \right)
\ee
The iteration matrix simplifies for large $r$ to
\smallskip\newline
\be
\left( \matrix{-\sqrt{[r]_q/[r+1]_q}
& 0 \cr
0 & - \sqrt{[r-1]_q/[r]_q}} \right)
\ee
and eventually goes like
\be
\left( \matrix{-1/{\vert}q\vert
& 0 \cr
0 & -1/{\vert}q{\vert} } \right)
\ee
Since $q^2 > 1$ we can thus apply the quotient criterium (behaviour
like a geometrical series) to conclude that
\be
\sum_{r=0}^{\infty} f_r^*(\lambda ) f_r(\lambda ) < \infty
\ee
i.e. that all $v_{\lambda}$ are
normalisable\footnote{It also implies that there are no normalisable
eigenvectors when $q^2 <1$}.
 They are obviously
contained in $D_{x^*}$ and are thus eigenvectors of $x^*$.
Since there are nonreal eigenvalues
 we conclude that $x^*$ is not symmetric.
This allows its eigenvectors to be linearily dependend. They are
actually in general linearily dependend on each other since the
Hilbert space $\cal{H}$ is seperable and there is an uncountable
infinite number of eigenvectors $v_{\lambda}$.
An analytic expression for the scalar product of two
normalised eigenvectors
$\langle {\hat{v}}_{\lambda},{\hat{v}}_{{\lambda}^{\prime}}\rangle $
has not yet been worked out. However, the numerical
approximation converges as
quickly as a geometrical series.
%
%
%
%
%
\vskip0.4truecm
The operator $x^{**}$ is much better behaved
than $x^*$, since it is closed and symmetric,
as every bi-adjoint of a densly defined
symmetric operator.
\smallskip\newline
Its domain
\begin{equation}
D_{x^{**}} = \left\{ v \in {\cal{H}} \vert \quad \exists w \in
{\cal{H}} \quad \forall a \in D_{x^*} :\quad
  \langle v,x^*.a\rangle = \langle w,a\rangle \right\}
\end{equation}
is in between those of $x$ and $x^*$:
 $D_x \subset D_{x^{**}} \subset D_{x^*}$
and it does
not contain any eigenvectors $v_{\lambda}$.
\subsection{Self-adjoint extensions}
We now apply the standard
procedure, see e.g. {\cite{hirze,akhi}}\footnote{Note that
\cite{hirze} defines 'hermitean'
as synonymous to self-adjoint while
\cite{akhi} uses it as synonymous to symmetric.},
for checking for
self-adjoint extensions of closed symmetric
operators:
\smallskip\newline
The idea is to check whether the 'Cayley transform' can be
isometrically extended.
An inverse Cayley transform then yields a self-adjoint
extension of $x^{**}$.
To this end we calculate the
orthogonal complement of the spaces:
\be
L_{\pm i,x^{**}} := (x^{**} \pm i).D_{x^{**}}
\ee
Since $x^{**}$ is closed, symmetric and has
$\bar{D}_{x^{**}} = \cal{H}$, these deficiency
subspaces $L_{\pm i,x^{**}}^{\perp}$ can be written as
\begin{equation}
L^{\perp}_{\pm i,x^{**}} := ker(x^* \mp i).D_{x^*}
\end{equation}
Here we used that $x^{***} = x^*$ which holds
because $x^*$ is closed and $\bar{D}_{x^*} = \cal{H}$.
Since there is only one $v_{i}$ and one $v_{-i}$ the dimensions of
these spaces, i.e. the deficiency indices are both equal to $1$.
We can thus define the following one-parameter family of
self-adjoint extensions:
\begin{equation}
x_{sa}(\phi).a := i (b + U.b) \mbox{\quad for all \quad }
a = b - U.b
\end{equation}
with the isometric operator $U$ defined on
$(x^{**}+i).D_{x^{**}} \oplus \mbox{{\bf C\rm}}v_{i}$ as
\begin{equation}
U.v := (x^{**} - i)(x^{**} +i)^{-1}.v \qquad
\forall v \in (x^{**}+i).D_{x^{**}} = L_{+i,x^{**}}
\end{equation}
and
\begin{equation}
U.\alpha v_i := \alpha e^{i \phi} v_{-i}
\end{equation}
 Here $\phi $ is a free real parameter,
labeling the self-adjoint extensions.
For the eigenvalues one can stay with
the extended Cayley transform $U$,
calculate its eigenvalues, and an
inverse M{\"o}bius transform then maps them
onto the eigenvalues of $x_{sa}(\phi )$.

The analysis for $p$ analogously leads to a one-parameter family
of self-adjoint extensions $p_{sa}(\psi)$.
One may now be tempted to
try to fix the choice of the extension
parameters $\phi $ and $\psi $
by requiring that $x_{sa}(\phi)$ and
 $p_{sa}(\psi)$ be defined on the
same domain. One would then like to
 diagonalise $x_{sa}(\phi)$
to obtain a coordinate space representation
or to diagonalise $p_{sa}(\psi)$
to obtain a momentum space representation.

However, we know from section 3 that $x$ and
$p$ cannot be extended to a
{\it common\rm }  domain on which they have eigenvectors.
\subsection{The $n$ dimensional case}
One finds in the $n$ dimensional case essentially the same situation:

We calculate the matrix elements of e.g. the position operator
in $j$-direction $x^{j}$ in the orthonormal basis of the vectors:
\be
e_{s_1,...,s_n} := \frac{(a_1^{\dagger})^{s_1} \cdot ... \cdot
 (a_n^{\dagger})^{s_n}}{\sqrt{[s_1]_q! \cdot ... \cdot [s_n]_q!}}
 \vert 0\rangle
\ee
Using Eqs.\ref{eins}-\ref{vier} one finds the matrix elements:
\begin{eqnarray}
x^{j}_{r_1,...,r_n,s_1,...,s_n} = & &
L_j (\sqrt{[r_j]_q} \delta_{r_j,s_j+1}
+ \sqrt{[s_j]_q} \delta_{r_j +1,s_j}) \cdot \nonumber \\
 & & q^{r_1+...+r_{j-1}} \delta_{r_1,s_1} \cdot ... \cdot
\delta_{r_{j-1},s_{j-1}}
\delta_{r_j+1,s_j+1}\cdot ... \cdot
\delta_{r_n,s_n}
\end{eqnarray}
The eigenvalue problem
\be
x_j v_{\lambda} = \lambda v_{\lambda}
\ee
is then solved by all
\be
v_{\lambda} = \sum_{s_j = 0}^{\infty} f_{s_1,...,s_j,...,s_n}(\lambda )
\quad e_{s_1,...,e_n}
\ee
with coefficients that obey the recursion formula (all $s_i$ are kept
fixed except $s_j$):
\begin{eqnarray}
\sqrt{[r_j+1]_q} f_{s_1,...,s_j+1,...,s_n} (\lambda ) & = &
\frac{\lambda }{L} q^{-(r_1+...+r_{j-1})} f_{s_1,...,s_j,...,s_n}
(\lambda )\nonumber \\
 &   & - \sqrt{[r_j]_q} f_{s_1,...,s_j-1,...,s_n} (\lambda )
\end{eqnarray}
The scaling factor $q^{-(r_1+...+r_{j-1})}$ implies that $x_j$ 'sees'
not only its own direction, reflecting that the position operators
no loeger commute (Eq.\ref{pospos}). Nevertheless
the same arguments as for the one dimensional case go through
again and we conclude that $v_{\lambda }$ is normalisable
for all complex $\lambda $. Again the
 operators $x_k^*$ are nonsymmetric
while the operators $x_k^{**}$ are symmetric and closed.
There is however a new feature which
is that the deficiency subspaces are now much larger than in the
one dimensional case because the operators $x_k^*$ now have countable
infinitely many linearly independend eigenvectors $v_i$ and $v_{-i}$.
Since both deficiency subspaces are
of the same size, $x_k$ has self-adjoint extensions.
The same arguments apply for the momentum operators $p_k$.
However, we conclude again from the uncertainty relation that there
is no common domain on which all these
operators are symmetric and have
eigenvectors.
\section{Summary and Outlook}
We thus arrive at the following picture:

While in classical mechanics the states can
have exact positions and momenta,
in quantum mechanics there is the
well known uncertainty principle, not
allowing $x$ and $p$ to have common
eigenvectors. Nevertheless
$x$ and $p$ seperately do have 'eigenvectors',
though non-normalisable ones. The
spectrum is continuous, namely the whole configuration or momentum
space.

{}From the above discussion we conclude that the 'noncommutative
geometry'- or quantum group generalisation of the Heisenberg algebra
has further consequences for the $x$ and $p$:
It is not only that the $x$ and $p$
have no {\it common eigenvectors\rm}, here they even have no
{\it common domain \rm }  on which they (are symmetric and) have
eigenvectors.
The $x$ and $p$ {\it seperately\rm} do have (infinitely many)
self-adjoint extensions and all candidates for eigenvectors are
normalisable. Thus the spectra of
the self-adjoint extensions of the position and the momentum operators
are no longer continuous. Only discrete ('lattices' of) eigenvalues
can occur.

In general one will however represent the $x$ and $p$, i.e. the
full Heisenberg algebra, on a common domain
 on which the $x$ and $p$ are symmetric. The symmetry is
of course to
insure that all physical expectation values are real.
We concluded that in this case the
$x$ and $p$ cannot have eigenvectors.
The non-existence of position or momentum eigenvectors
meant of course the non-existence of absolute precision in
position or momentum measurements. We found
 minimal nonzero uncertainties in these
measurements.
The maximal common
domain on which the $x$ and $p$ are symmetric
remains to be determined.

It should be interesting to examine whether this
quantum mechanical formalism with
 'built in' minimal uncertainties in the $x$ and $p$
can find applications in effective
theories, where minimal nonzero uncertainties
in position or momentum measurements appear naturally,
e.g. in solid state or nuclear physics.

The present paper also
supports the idea that noncommtutive geometry mathematics
has the potential to regularise the small $\Delta x$ i.e. the
ultraviolet, as well as the small $\Delta p$, i.e. the infrared
behaviour of quantum field theories.
\medskip\newline
\bf Acknowledgements\rm \newline
I would like to thank J.Mickelson and J.Wess for their interest and
useful critisims.


\begin{thebibliography}{10}

\bibitem{DRI} V.G.Drinfel'd, in Proc. ICM Berkeley,
AMS, Vol. 1 (1986) 798-820
\bibitem{FRT} L.D.Faddeev, N.Yu.Reshetikhin, L.A.Takhtajan,
Alg.Anal. 1, 1, (1989) 178
\bibitem{MAJ1} S.Majid, Int. J. Mod. Phys. A. Vol. 5, No 1, (1990) 1-91
\bibitem{Connes} A.Connes, Publ. I.H.E.S. 62 (1986) 257
\bibitem{ixtapa} A.Kempf, to appear in Proc. XXII DGM Conference, Sept. '93,
Ixtapa, Mexiko
\bibitem{meinslmp} A.Kempf, Lett. Math. Phys. 26: (1992) 1-12
\bibitem{woro} W.Pusz, S.Woronowicz, Rep. Math. Phys. 27 (1989) 231
\bibitem{dcwess} J.Wess, B.Zumino, Nucl. Phys. Proc. Suppl. 18 B (1991) 302
\bibitem{MAJ2} S.Majid, J. Math. Phys. 34 (10) (1993) 4843-4856
\bibitem{macf} A.Macfarlane, J. Phys. A 22 (1989) 4581
\bibitem{bieden} L.Biedenharn, J. Phys. A 22 (1989) L 873
\bibitem{chef} J.Schwenck, J.Wess, Phys. Lett. B 291, (1992) 273
\bibitem{arthur} A.Hebecker, W.Weich, in Proc. XIX Symposium Ahrenshoop,
Eds. B.D{\"o}rfel, E.Wieczorek, DESY Preprint 93-013, (1992)
\bibitem{akny} A.Kempf, in Proc. XX DGM- Conference, June '91, New York,
Eds. S.Catto, A.Rocha, World Scientific, (1991) 546
\bibitem{berlin} A.Kempf, in Proc. XIX Symposium
Ahrenshoop in Wendisch-Rietz,
Eds. B.D{\"o}rfel, E.Wieczorek, DESY Preprint 93-013, (1992)
\bibitem{baulieu} L.Baulieu, E.G.Floratos, Phys. Lett. B 258 (1991) 171
\bibitem{meinsjmp} A.Kempf, J. Math. Phys., Vol.
34, No.3, (1993) 969-987
\bibitem{hirze} F.Hirzebruch, W.Scharlau, 'Einf. Funktionalanalysis',
B.I. (1971)
\bibitem{akhi} N.I. Akhiezer, I.M. Glazman, 'Theory of Lin. Op. in
Hilbert space', F. Ungar Publ. (1963)
\end{thebibliography}
\end{document}